\documentclass[prb,twocolumn,showpacs,superscriptaddress,preprintnumbers,floatfix,amsmath,amssymb]{revtex4-1}
\usepackage{amssymb}
\usepackage{graphicx}% Include figure files
\usepackage{dcolumn}% Align table columns on decimal point
\usepackage{bm}% bold math
\begin{document}

\title{Multiband superconductivity in Ta$_4$Pd$_3$Te$_{16}$ with anisotropic gap structure}

\author{Wen-He Jiao} \email{whjiao@zust.edu.cn}
\affiliation{School of Science, Zhejiang University of Science and Technology, Hangzhou 310023, China}

\author{Yi Liu}
\affiliation{Department of Physics, Zhejiang University, Hangzhou
310027, China}

\author{Yu-Ke Li}
\affiliation{Department of Physics, Hangzhou Normal University, Hangzhou 310036, China}

\author{Xiao-Feng Xu}
\affiliation{Department of Physics, Hangzhou Normal University, Hangzhou 310036, China}

\author{Jin-Ke Bao}
\affiliation{Department of Physics, Zhejiang University, Hangzhou
310027, China}

\author{Chun-Mu Feng}
\affiliation{Department of Physics, Zhejiang University, Hangzhou
310027, China}

\author{S. Y. Li}
\affiliation{State Key Laboratory of Surface Physics, Department of Physics, and Laboratory of Advanced Materials, Fudan University, Shanghai 200433, China}
\affiliation{Collaborative Innovation Centre of Advanced Microstructures, Nanjing University, Nanjing 210093, China}

\author{Zhu-An Xu}
\affiliation{Department of Physics, Zhejiang University, Hangzhou
310027, China} \affiliation{Collaborative Innovation Centre of Advanced Microstructures, Nanjing University, Nanjing 210093, China}

\author{Guang-Han Cao} \email{ghcao@zju.edu.cn}
\affiliation{Department of Physics, Zhejiang University, Hangzhou
310027, China} \affiliation{Collaborative Innovation Centre of Advanced Microstructures, Nanjing University, Nanjing 210093, China}

\date{\today}

\begin{abstract}
We carried out the measurements of magnetoresistance, magnetic susceptibility and specific heat
on crystals of the low-dimensional transition metal telluride
Ta$_4$Pd$_3$Te$_{16}$. Our results indicate that Ta$_4$Pd$_3$Te$_{16}$ is an
anisotropic type-II superconductor with the extracted Ginzburg-Landau parameter
$\kappa_{\text{GL}}=$ 84. The upper critical field $H_{c2}$($T$) shows a linear dependence
at low temperature and the anisotropy of $H_{c2}$($T$) is strongly $T$-dependent,
both of which indicate a multiband scenario. A detailed analysis reveals that the
electronic specific heat $C_{\text{el}}$($T$) can be consistently described by a
two-gap ($s$+$d$ waves) model from the base temperature $T/T_c\sim$ 0.12 up to $T_c$.
Our data suggests multiband superconductivity in Ta$_4$Pd$_3$Te$_{16}$ with anisotropic gap structure.
\end{abstract}


%74.70.-b Superconducting materials other than cuprates
%74.78.-w Superconducting films and low-dimensional structures
%74.25.Op Mixed states, critical fields, and surface sheaths
%74.25.Bt Thermodynamic properties
\maketitle

Exploring exotic pairing mechanisms in new superconducting compounds have
always been one of the most attractive issues in condensed matter physics.
Although the Bardeen-Cooper-Schrieffer (BCS) theory\cite{BCS} provides excellent
explanations for the conventional phonon-mediated $s$-wave superconductor, the
successive emergence of unconventional superconductivity, claimed to be in either
the singlet $s_{\pm}$-wave\cite{s} and $d$-wave pairing states,\cite{d1,d2} or the
triplet $p$-wave\cite{p1,p2} and $f$-wave pairing states,\cite{f} presents great
challenges for physicists to uncover the variously exotic non-phonon-mediated
pairing mechanisms. The spin or charge fluctuations, which always appear in
low-dimensional structures, were often regarded as the pairing glues for a large number
of unconventional superconductors, such as the cuprates,\cite{Lee-d} iron-based
superconductors,\cite{s} quasi-one-dimensional (Q1D) superconductors,\cite{Organic-1,Organic-2}
and heavy-fermion superconductors.\cite{CePd2Si2,UPd2Al3} Among them, multiband
superconductivity and symmetry-imposed nodes were commonly observed as evidenced
by a variety of techniques including thermodynamic approaches.\cite{BaKFeAs,FeTeSe,TlNiSe}
In this regard, the gained information on the gap symmetry has significant meanings
in understanding the novel pairing mechanisms.

Recently, we discovered bulk superconductivity with $T_c$ = 4.6 K in a transition metal telluride Ta$_4$Pd$_3$Te$_{16}$.\cite{Ta4316-J} This material has a layered structure with Q1D characteristics
consisting of PdTe$_2$ chains, TaTe$_3$ chains and Ta$_2$Te$_4$ double chains along the
crystallographic $b$-axis. The ($\bar{1}$03) planes, as illustrated in the inset of Fig. 1(c), are formed
by the connection of the chains mentioned above along the [301] direction. These structural features
are well reflected in the morphology of the crystals as shown in the inset of Fig. 1(a). For simplicity,
hereafter we define $a^*$-axis as to be parallel to the [301] direction and $c^*$-axis as to be
perpendicular to the ($\bar{1}$03) plane. Compared with layered chalcogenide superconductors Nb$_3$Pd$_x$Se$_7$\cite{Nb3Pd0.7Se7} and $M$$_2$Pd$_x$$Q$$_5$ ($M$ = Nb and Ta, $Q$ = S and Se)\cite{Nd2Pd0.81S5,Nb2PdxSe5,Ta2PdxS5,Nb2PdS5} with similar
Q1D structures, Ta$_4$Pd$_3$Te$_{16}$ is a stoichiometric compound with flat two-dimensional sheets and
it shows stronger electron-electron interactions.\cite{Ta4316-J} Thermal conductivity measurements
demonstrate the presence of nodes in the superconducting gap by the evidences of a large residual
$\kappa_0/T$ in zero field and the rapid increase of $\kappa_0/T$ in magnetic field, mimicking the
behavior of $d$-wave cuprate superconductor Tl-2201.\cite{TT-Li} Furthermore, a temperature-pressure superconducting
dome was as well observed, leading to the expectation that certain density-wave orders may neighbour the
superconducting dome. According to the result of electronic structure calculations made
by Singh,\cite{Singh} any nearby density wave instability, if exists, would be a charge-density-wave
(CDW) but not a spin-density-wave (SDW) as the transition metal contribution to the density of states
is rather small. Besides, an $s$-wave pairing state from phonons associated with the Te-Te $p$ bonding
and a Fermi surface associated Te $p$ bands are argued to be the most possible case. To explain
the results of the thermal conductivity, the author also proposes the likelyhood of Ta$_4$Pd$_3$Te$_{16}$
being a clean multi-gap superconductor where the gap ratio is significantly large.\cite{Singh} However, more recently,
the low-temperature scanning tunneling spectroscopy (STS) measurements show the evidence for the
presences of anisotropic superconducting gap with gap minima or even node.\cite{Ta4316-whh}
Thus, the issue of the gap symmetry in Ta$_4$Pd$_3$Te$_{16}$ is worthy to be addressed further.

\begin{figure*}
\includegraphics[width=8.9cm]{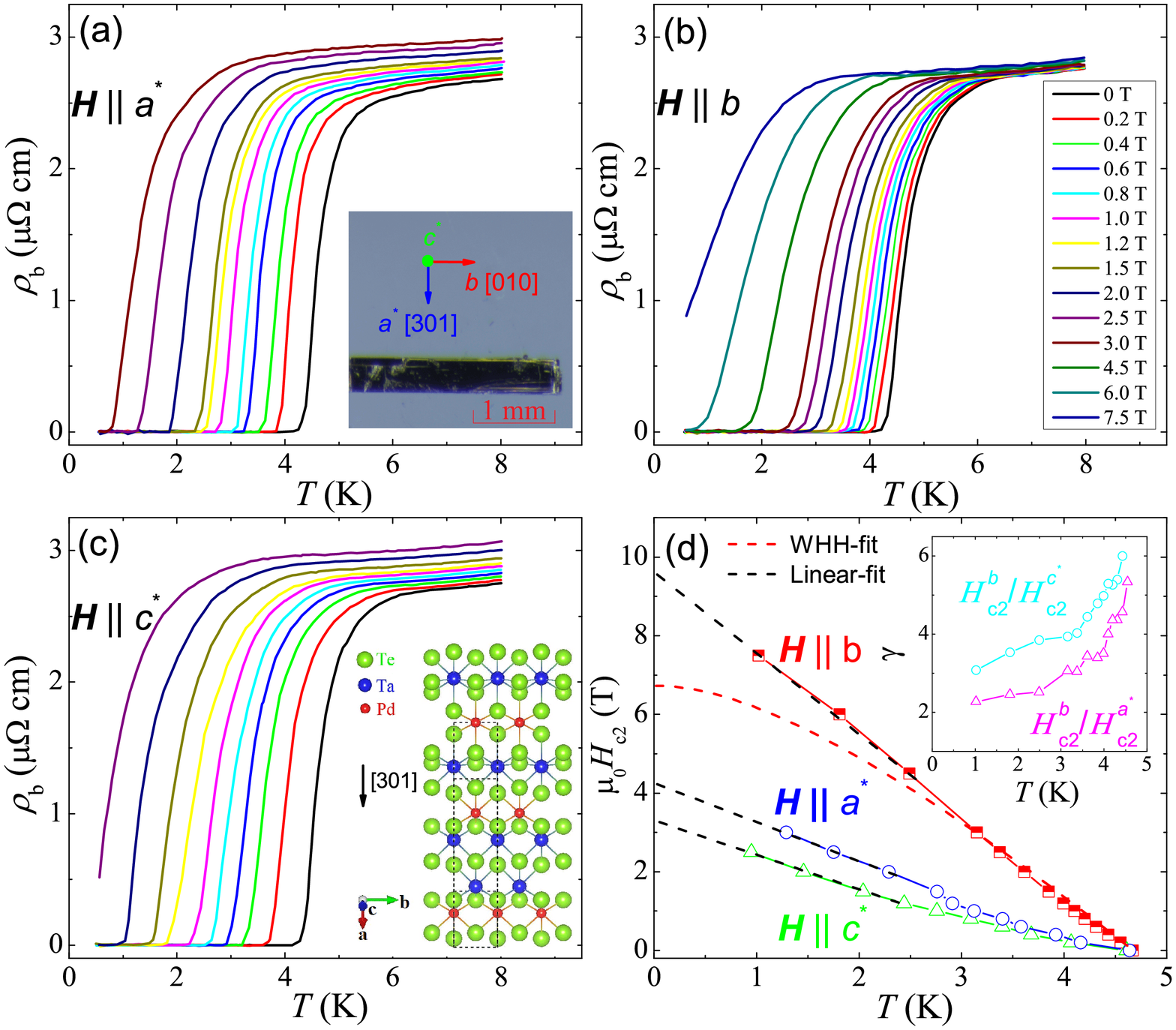}
\caption{(Color online) Magnetoresistivity of the Ta$_4$Pd$_3$Te$_{16}$ crystal for (a)
$\emph{\textbf{H}}\parallel a^{*}$, (b) $\emph{\textbf{H}}\parallel b$ and (c)
$\emph{\textbf{H}}\parallel c^{*}$. (d) The extracted superconducting upper critical field
$H_{c2}(T)$ for different field orientations. Werthamer-Helfand-Hohenberg
(WHH) fitting is made to the data for $\emph{\textbf{H}}\parallel b$, represented by the red
dashed lines. The black dashed lines show linear fittings. The inset of (a) shows the image of
the crystal under an optical microscope, from which the crystallographic directions can be
easily identified. The crystal structure of Ta$_4$Pd$_3$Te$_{16}$ viewed perpendicular to the ($\bar{1}$03)
plane is plotted in the inset of (c). The inset of (d) shows the anisotropy of $H_{c2}(T)$.}
\end{figure*}

In this Rapid Communication, we investigate the superconducting properties of the Ta$_4$Pd$_3$Te$_{16}$
crystals via systematic magnetoresistivity (MR), magnetic susceptibility and specific heat measurements.
The results indicate that Ta$_4$Pd$_3$Te$_{16}$ is an anisotropic type-II superconductor. The upper
critical fields $H_{c2}$(0) were estimated to be 4.2, 9.6 and 3.3 T for fields applied along the three
axes, $a^{*}$, $b$ and $c^{*}$, respectively. We observed strong $T$-dependent anisotropy of
$H_{c2}$($T$) and the linear increase $H_{c2}$, both of which suggest the multiband superconductivity.
The zero-field heat capacity data can be best described by a two-gap model with anisotropic gap structure.
In addition,  the electronic specific heat coefficient in the mixed state, $\gamma_e(H)$, exhibits a
nonlinear behavior. Thus, our results suggest two energy gaps with anisotropic gap structure are associated
with the superconductivity in Ta$_4$Pd$_3$Te$_{16}$.

Single crystals of Ta$_4$Pd$_3$Te$_{16}$ were grown by a self-flux technique as previously
described.\cite{Ta4316-J} High quality of as-grown crystals was confirmed by X-ray
diffraction. To insure the data obtained more reliable, we elaborately selected crystals for
the measurements, all of which having shiny surfaces without any discernible solvent attached.
MR measurements were carried out by a stand four-probe technique with current $I=$ 2 mA applied
along the $b$-axis. The specific heat for a collection of five needle-like crystals with a total
mass $m$ = 4.68(2) mg was measured by a long relaxation method utilizing a commercial $^3$He
microcalorimeter (Quantum Design PPMS-9). In various magnetic fields, the thermometer on the
calorimeter puck was calibrated before the measurements, and the addenda was predetermined in a
separate run. Magnetic susceptibility measurement was performed using a superconducting quantum
interference device magnetometer (MPMS-5).

Figure 1 encapsulates the $b$-axis MR ($\rho_b$) and the extracted $H_{c2}$ with fields
applied along $a^{*}$, $b$ and $c^{*}$, respectively. To eliminate the ambiguity from the
superconducting fluctuations, the 50\% criterion was used in determining $H_{c2}$,
\emph{i.e.}, the field at which $\rho_b$ reaches 50\% of the normal state resistivity.
We employed the Werthamer-Helfand-Hohenberg (WHH) formula for an isotropic one-band BCS
superconductor in a dirty limit to fit $H_{c2}^b(T)$.\cite{WHH} Apparently, at low
temperatures, the one-band WHH model fails to satisfy the extracted $H_{c2}^b(T)$.
By linear extrapolations, $H_{c2}^{a^{*}}$, $H_{c2}^b(T)$ and $H_{c2}^{c^{*}}$ are
estimated to be 4.2, 9.6 and 3.3 T, respectively. The estimated value of $H_{c2}^b(T)$
is close to the Pauli-Clogston limiting field $H_p=1.84T_c\sim8.5$ T, excluding the
possibility of triplet pairing. Besides, all of the three curves show
a linear $T$-dependence at low temperatures, which is consistent with the previous report\cite{TT-Li}
but in sharp contrast with what it should display in most superconductors,\cite{WHH,Tinkham} \emph{i.e.},
a concave down curvature at low temperatures. This anomalous behavior was regarded as the
evidence for multiband superconductivity in iron-based superconductors.\cite{Yuan,Kim}
Moreover, as shown in the inset of Fig. 1(d), the strong $T$-dependent superconducting anisotropy
$\gamma=H_{c2}^b/H_{c2}^{c^{*}}$ or $H_{c2}^b/H_{c2}^{a^{*}}$ provides
further evidence for multiband scenario as the case in two-band superconductor MgB$_2$ and most
iron-based superconductors.\cite{MgB2,FeSC} Using the anisotropic GL formula
\begin{equation}
H_{c2}^i(0)=\phi_0/2\pi\xi_j(0)\xi_k(0),
\end{equation}
where $\phi_0$ is the flux quantum and $\xi_j$(0) is the GL coherence length along the $j$-direction,
$\xi(0)$ at zero temperature is calculated to be 66.1, 151.0 and 51.9 $\text{{\AA}}$ for $a^{*}$, $b$
and $c^{*}$, respectively. Obviously, the interchain coherence lengths $\xi^{a^{*}}$(0) and $\xi^{c^{*}}$(0)
are much longer than the distance between any two adjacent chains in Ta$_4$Pd$_3$Te$_{16}$ denoting the
continuous superconducting phase across the chains. In other words, the superconductivity here is anisotropic
but three-dimensional in nature, similar to recent discovered Q1D superconductors Nb$_2$Pd$_x$Se$_5$ and Ta$_2$Pd$_x$S$_5$.\cite{Nb2PdxSe5,Ta2PdxS5} For the normal state, the longitudinal MR ($\emph{\textbf{H}}\parallel\emph{\textbf{I}}$) response is negligibly small, while the in-chain MR ($\emph{\textbf{H}}\perp\emph{\textbf{I}}$) is obviously large in magnitude and positive, which is typical
of a Q1D metal due to the change in carrier trajectories induced by the Lorentz force.\cite{MR-PrBaCuO}

\begin{figure}
\includegraphics[width=5.5cm]{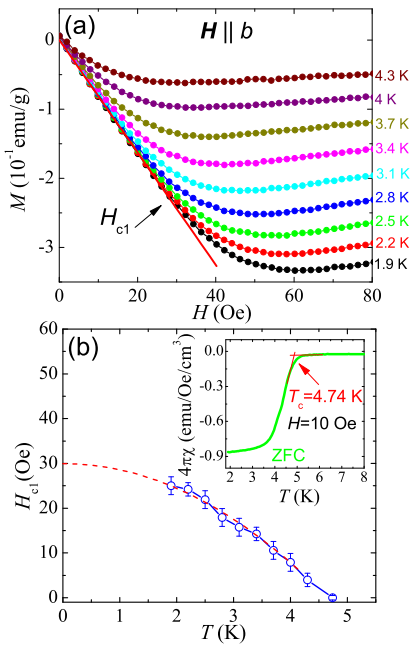}
\caption{(Color online) (a) The low-field magnetization curve $M(H)$
under field applied along $b$-axis at various temperatures. The solid
red line is fitted to the 1.9 K data. (b) The extracted $H_{c1}^{b}(T)$
for Ta$_4$Pd$_3$Te$_{16}$. The inset shows the diamagnetic signal under
10 Oe with zero-field-cooling (ZFC) mode.}
\end{figure}

The lower critical field values, $H_{c1}^{b}(T)$, were determined from low-field magnetization curves
$M(H)$ with field applied along the $b$-axis as shown in Fig. 2. The low-field parts almost overlap
with the Meissner line (the solid red line in Fig. 2(a)) due to the Meissner effect. Thus,
$H_{c1}^{b}(T)$ could be defined at the point where $M(H)$ deviate by 2\% from the perfect Meissner
response, and the extracted results are plotted in Fig. 2(b). Since that our crystal is needle-like
and the demagnetization factor for field along the needle-like direction is negligibly small,
$H_{c1}^{b}$(0) can be directly estimated to be 29.9 Oe by fitting the extracted data to the formula, $H_{c1}(T)=H_{c1}(0)[1-(T/T_c)^2]$,
represented by the dashed red line in Fig. 2(b). With the results for $H_{c1}^{b}(0)$ and $H_{c2}^{b}(0)$,
we calculated the GL parameter $\kappa_{\text{GL}}^b$ to be about 84 using the equation $H_{c2}(0)/H_{c1}(0)=2\kappa_{\text{GL}}^2/\text{ln}\kappa_{\text{GL}}$.
The above results indicate that Ta$_4$Pd$_3$Te$_{16}$ is an extremely type-II superconductor.

Figure 3 presents the $T$-dependent specific heat of Ta$_4$Pd$_3$Te$_{16}$ crystals
divided by temperature at zero magnetic field. A sharp anomaly, denoting the superconducting
transition, can be clearly observed at $T_c\sim$ 4.14 K (defined by an entropy conserving construction).
In Fig. 3(a), the raw heat capacity data between 3.5 K and 7 K is fitted to the formula
$C/T=\gamma+\beta T^2$, and the fitting gives parameters $\gamma=$ 51.2(1) mJ/mol K$^2$ and
$\beta=$ 13.53(1) mJ/mol K$^4$, both of which are slightly larger than our previous report.\cite{Ta4316-J}
This small discrepancy is possibly due to the measurement precision as well as the uncertainty
of sample mass weighed. Besides, a small residual linear term $\gamma_0=$ 5.1 mJ/mol K$^2$ exists,
as observed from the intercept of $C/T$ vs $T^2$ in the $T\rightarrow 0$ limit, the origin
of which may be the presence of nodal quasiparticles for a nodal superconducting gap or
a small fraction of nonsuperconducting metallic impurity phase.
By subtracting the phononic contribution, the electronic specific heat can be obtained by:
$C_{\text{el}}(T)=C(T)-\beta T^3$, which is plotted in Fig. 3(b) as
$C_{\text{el}}/T$ versus $T/T_c$. The zero-field data at low temperature is obviously
irreconcilable with the conventional $s$-wave order parameter as the significant
quasi-particle excitations.

\begin{figure}
\includegraphics[width=8.5cm]{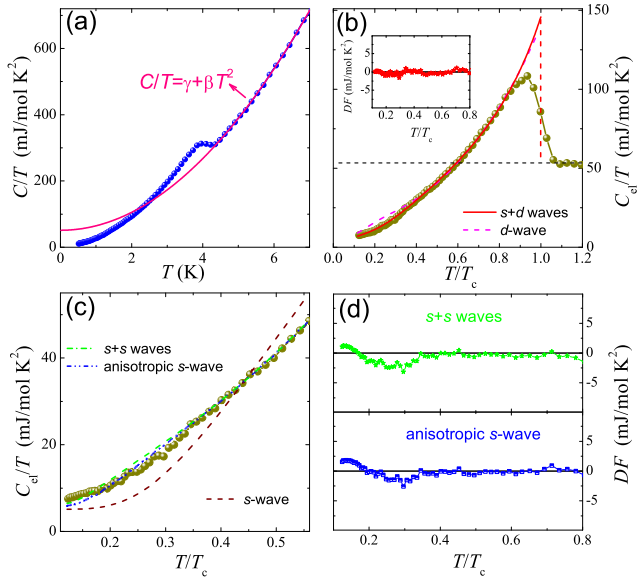}
\caption{(Color online) (a) Specific heat of Ta$_4$Pd$_3$Te$_{16}$ crystals
divided by temperature ($C/T$) at zero magnetic field. The solid pink line is
the fit to the normal-state data from 4.5 to 7 K using the formula $C/T=\gamma+\beta T^2$.
(b) $T$-dependent electronic contribution to the specific heat divided
by temperature ($C_{\text{el}}/T$). Red and magenta dashed lines are the fitting curves with
two-gap ($s$+$d$ waves) and single-gap $d$-wave scenarios. The inset shows the deviation of the fitting
value for the case of $s$+$d$ waves with the data (represented by DF in the plot).
(c) expands the low temperature region. Lines show the theoretical fitting of various $s$-wave gap
functions as described in the text. (d) The deviation of the fitting from the data for the cases of $s$+$s$ waves
(upper panel) and anisotropic $s$-wave (lower panel).}
\end{figure}

The so-called $\alpha$ model, which was devised to simulate the strong-coupling
effect, has gained significant success in explaining the properties of strong-coupling
superconductors.\cite{alfa} Within this model, the temperature dependence of
energy gap $\Delta(T)$ approximately follows weak-coupling BCS behavior multiplied
by a dimensionless parameter $\alpha$. In the BCS theory, the entropy $S$ in the superconducting
state is given by\cite{Bouquet}
\begin{equation}
S=-\frac{3\gamma_n}{k_B \pi^3}\int_{0}^{2\pi}\int_{0}^{\infty}f\textrm{ln}f+
[1-f]\textrm{ln}[1-f]d\varepsilon d\phi,
\end{equation}
where $f$ is the Fermi function $f=(1+e^{E/k_BT})^{-1}$ with the
quasiparticle energy $E=\sqrt{\varepsilon^2+\Delta^2(T,\phi)}$, $\gamma_n$ is
the normal state $\gamma$ for the superconducting part, and $\Delta(T,\phi)$
is the temperature and angle dependence of the gap function. The electronic heat
capacity is thus calculated by $C_{\text{el}}=T(\partial S/ \partial T)$.
To elucidate the superconducting order parameter, the raw data of $C_{\text{el}}$
is analyzed and fitted by various models, $i.e.$, the isotropic $s$-wave function,
anisotropic $d$-wave function, two-fold-symmetric anisotropic $s$-wave function [$\Delta(\phi)=\alpha_1+\alpha_2\cos{2\phi}$], and two-gap scenarios
($s$+$s$ waves and $s$+$d$ waves). The contribution of $C_{\text{el}}$ from
nodal quasiparticles or the small fraction of nonsuperconducting metallic impurity phase
is regarded as $\gamma_0 T$ in our fittings. The fitting curves for the cases of single-band
$d$-wave and two-band $s$+$d$ waves are plotted in Fig. 3(b), from which,
one can unambiguously observe that the two-gap ($s$+$d$ waves) model best captures the
experimental data from the base temperature $T/T_c\sim$ 0.12 up to $T_c$.
The fitting gives the following parameters: $\alpha^s=$ 0.916 (for $s$ wave),
$\alpha^d=$ 1.652 (for $d$ wave) and the ratio of weight $\gamma_n^s:\gamma_n^d=$ 33.5 $:$ 66.5,
where the partial Sommerfild coefficient $\gamma_n^{s(d)}$ characterizes each band.
Nodal quasiparticles for the nodal superconducting gap or impure phase are expected to
be the source of the finite value $\gamma_0 $ in $T\rightarrow 0$ limit for this case.
The fittings with other gap functions (full-gap cases) mentioned above deviate the experimental data
significantly, especially in low-temperature region, the close-up view of which is
displayed in Fig. 3(c). For the full-gap cases, the residual linear term $\gamma_0 $
should come from a small amount of nonsuperconducting metallic impure phase. In
order to show more clearly the deviations of the fitting curves from the data for the
case of $s$+$s$ waves and anisotropic $s$-wave, we plots the differences in Fig. 3(d).
Therefore, our result strongly suggests Ta$_4$Pd$_3$Te$_{16}$ is a two-gap superconductor
with the pairing symmetry of $s$+$d$ waves. This fitting is consistent with the STS study
as the temperature dependence of tunneling spectra can be well fitted with $s$+$d$ waves
(the anisotropic $s$-wave symmetry can as well describe their data while in our
heat-capacity study it can not).\cite{Ta4316-whh} In addition, by calculating the
maximum gap value using the derived parameters
($\Delta_{\text{max}}$ = 0.335$*\Delta^s$ + 0.665$*\Delta^d$), we find that
$\Delta_{\text{max}}/k_\text{B} T_c\approx$ 2.47, very close to the value 2.3 reported
in Ref. 24. The scenario of two-band $s$-wave gaps with significant ratio has been
proposed to be possible in understanding the data of thermal conductivity.\cite{Singh}
However, our analysis of the heat capacity does not provide sufficient evidence for this possibility.
On the contrary, the evidence of anisotropic multi-gap structure is consistent with the
study of thermal conductivity,\cite{TT-Li} which suggests the presence of node in the superconducting gap.
Overall, our result reported here reconciles the recent experimental studies regarding
the pairing symmetry in Ta$_4$Pd$_3$Te$_{16}$.

\begin{figure}
\includegraphics[width=6.65cm]{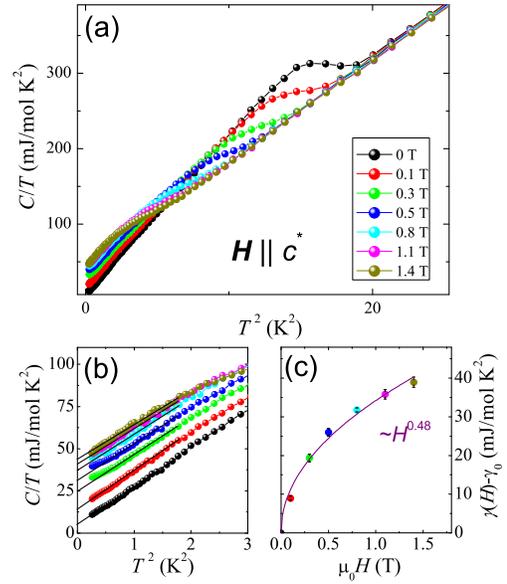}
\caption{(Color online) (a) Specific heat of Ta$_4$Pd$_3$Te$_{16}$ crystals in selected
magnetic fields parallel to $c^*$-axis, plotted as $C/T$ vs $T^2$. (b) The plots of $C/T$
vs $T^2$ in the low-$T$ region. For each applied field, $\gamma(H)\equiv$ lim$_{T\rightarrow0}$
$C(T,H)/T$ was determined from a linear extrapolation of the experimental data in
the temperature range 0.5$-$1.3 K. (c) Magnetic field dependence of the electronic specific
heat coefficient $\gamma(H)$ in the mixed state. }
\end{figure}

As the magnetic field dependence of specific heat at low temperatures is
instructive for the quasi-particle excitations, we measured the low-$T$
specific heat of Ta$_4$Pd$_3$Te$_{16}$ crystals under magnetic fields ($\mu_0 H\leq1.4$ T)
for $\emph{\textbf{H}}\parallel c^{*}$ as shown in Fig. 4(a). The heat capacity anomaly is
gradually suppressed to lower temperatures with the increase of field. The enlarged part of the
data is displayed in Fig. 4(b), from which, one can observe that minor upturns
under $\mu_0H=$ 0.3 and 0.5 T appear at low temperatures, probably stemming from Schottky
anomaly. Therefore, we extracted field-dependent electronic coefficient $\gamma(H)$ by
a linear extrapolation of $\gamma(H)\equiv$ lim$_{T\rightarrow0}$$C(T,H)/T$, without
considering the upturns under 0.3 and 0.5 T.
The obtained results are presented in Fig. 4(c) as $\gamma_e(H)=$ $\gamma(H)-\gamma_0$ vs $H$.
In fully gapped conventional type-II superconductors, it is generally known that $\gamma_e(H)$
should be linearly proportional to the number of field-induced vortices,\cite{s-wave}
\emph{i.e.}, $\gamma_e(H)\propto H$. However, we find in Ta$_4$Pd$_3$Te$_{16}$, a power-law
fitting provides the relation $\gamma_e(H)\sim H^{0.48}$ as shown in Fig. 4(c), thereby
ruling out the isotropic $s$-wave pairing symmetry.
In the case of two isotropic $s$-wave superconducting gaps, $\gamma_e(H)$ is featured by an
intersection of two straight lines, as observed in multiband superconductivity with two $s$-wave
gaps, \emph{e.g.}, MgB$_2$,\cite{MgB} NbSe$_2$,\cite{NbSe} and SrPt$_2$As$_2$.\cite{SrPtAs}
The nonlinear behavior observed here is seemingly in contrast with the above
scenario. Although $\sqrt{H}$ behavior was once considered as the evidence of $d$-wave superconductors
due to the Doppler shift of extended quasi-particle excitation spectrum,\cite{H0.5}
this kind of nonlinear behavior was also frequently observed in many
multiband superconductors, such as TlNi$_2$Se$_2$.\cite{TlNiSe} Overall, the nonlinear behavior
for $\gamma_e(H)$ could be regarded as an evidence for multiband superconductivity.

In summary, we have investigated the superconducting properties of Ta$_4$Pd$_3$Te$_{16}$ crystals.
Our measurements indicate Ta$_4$Pd$_3$Te$_{16}$ is a new anisotropic type-II superconductor
with the extracted Ginzburg-Landau parameter $\kappa_{\text{GL}}=$ 84.
The linear $T$-dependent $H_{c2}$ at low temperatures
and the strong $T$-dependent anisotropy of $H_{c2}$ suggest the multiband
superconductivity. Indeed, the zero-field electronic heat capacity data, $C_{\text{el}}$, can
be satisfactorily described in terms of a two-gap ($s$+$d$ waves) model.
The electronic heat capacity coefficient, $\gamma_e(H)$, exhibiting
a nonlinear behavior. Thus, Ta$_4$Pd$_3$Te$_{16}$
appears to be a rare example of a two-gap superconductor with the gap symmetry of $s$+$d$ waves rather
than a multiple fully gapped $s$-wave symmetry.
This superconducting state is very interesting and worthy to investigate further, especially when noting the Q1D structural characteristics in Ta$_4$Pd$_3$Te$_{16}$.

%\newpage
\begin{acknowledgments}
We would like to thank Huiqiu Yuan and Yongkang Luo for helpful
discussions. This work is supported by National Basic Research
Program of China (Nos. 2011CBA00103 and 2010CB923003) and National
Science Foundation of China (No. 11190023).
\end{acknowledgments}

%References

\end{document}